\documentclass[pra,aps,reprint]{revtex4-1}

\usepackage{etoolbox}
\usepackage{hyperref}
\usepackage{amsmath}
\usepackage{amssymb}
\usepackage{dsfont}
\usepackage[T1]{fontenc}
\usepackage{ae}
\usepackage{color}
\usepackage{graphicx} 
\usepackage{placeins}

\usepackage[mediumspace,mediumqspace,squaren]{SIunits} 

\begin{document}

\title{Manipulating cold atoms through a high-resolution compact system \\ based on a multimode fiber}

\author{Nicolas Vitrant} \affiliation{JEIP,  USR 3573 CNRS, Coll{\`e}ge de France, PSL University, 11, place Marcelin Berthelot, 75231 Paris Cedex 05, France}

\author{Kilian M{\"u}ller}  \email{Present address: LightOn, 6 Rue Jean Calvin, 75005 Paris, France} \affiliation{JEIP,  USR 3573 CNRS, Coll{\`e}ge de France, PSL University, 11, place Marcelin Berthelot, 75231 Paris Cedex 05, France}

\author{S{\'e}bastien Garcia} \affiliation{JEIP,  USR 3573 CNRS, Coll{\`e}ge de France, PSL University, 11, place Marcelin Berthelot, 75231 Paris Cedex 05, France}

\author{Alexei Ourjoumtsev} \email{Corresponding author: alexei.ourjoumtsev@college-de-france.fr} \affiliation{JEIP,  USR 3573 CNRS, Coll{\`e}ge de France, PSL University, 11, place Marcelin Berthelot, 75231 Paris Cedex 05, France}

\begin{abstract}
 
To manipulate cold atoms in spatially constrained quantum engineering platforms, we developed a lensless optical system with a $\sim$\unit{1}{\micro\meter} resolution and a transverse size of only \unit{225}{\micro\meter}. We use a multimode optical fiber with a high numerical aperture, which directly guides light inside the ultra-high-vacuum system. Spatial light modulators allow us to generate control beams at the in-vacuum fiber end by digital optical phase conjugation. As a demonstration, we use this system to optically transport cold atoms towards the in-vacuum fiber end, to load them in optical microtraps and to re-cool them in optical molasses. 
This work opens new perspectives for setups combining cold atoms with other optical, electronic or opto-mechanical systems with limited optical access.

\end{abstract}

\maketitle

\section{Introduction}

Cold atoms play a crucial role in the rapidly growing field of quantum engineering. Ever more demanding experiments on quantum logic, quantum measurements or quantum simulations require to image and optically control atoms with a high resolution. As standard microscope objectives are not compatible with ultra-high-vacuum environments required for atomic laser cooling, current experiments use diffraction-limited aspheres \cite{Sortais2007} or specially designed objectives \cite{Bakr2009,Sherson2010} placed in the direct vicinity of the atoms. However, many ``hybrid'' platforms also require to interface cold atoms with other devices such as optical cavities \cite{Thompson1992}, superconducting circuits \cite{Bernon2013} or mechanical resonators \cite{Hunger2011}. These devices restrict the optical access and make the use of bulk high-resolution optics impractical if not impossible, severely limiting the control over the atomic part of the system. 

Optical fibers, seen as ultra-compact flexible light guides, appear as a viable solution to this problem. Singlemode fibers, with \cite{Garcia2013} or without \cite{Vetsch2010} additional optics, have been used in this context, but their operation remains restricted to pre-determined optical configurations. In contrast, high numerical aperture multimode fibers are lensless systems allowing one to create arbitrary reconfigurable light patterns with a $\sim$\unit{1}{\micro\meter} optical resolution~\cite{Papadopoulos2012}, provided that one can reshape the strongly distorted transmitted optical wavefronts with adaptive optics~\cite{Popoff2010}.  

In this letter, inspired by methods developed for biomedical imaging~\cite{Yaqoob2008,Cui2010,Papadopoulos2013}, we use beams propagating through a multimode optical fiber to manipulate a cold atomic ensemble. The control beams are generated by digital optical phase conjugation (DOPC) with spatial light modulators (SLM). By using far-detuned laser beams, we transport atoms in an optical lattice and we transfer them to optical microtraps at $200\,\mu$m from the fiber end face. With near-resonant light fields, we generate optical molasses to Doppler cool the atoms in front of the fiber. Our approach proves to be compatible with the typical constrains of cold-atom setups and thus constitutes a new compact and versatile tool for such experiments. 

\section{Experimental setup}
\label{SectSetup}

The core of our optical system is a commercial \unit{25}{\centi\meter} long multimode optical fiber (MMF, Thorlabs FP200ERT) with a numerical aperture NA=$0.5$, a core diameter of \unit{200}{\micro\meter} and a cladding diameter of \unit{225}{\micro\meter}. The ``distal'' end of this fiber, stripped of its Tefzel coating, is permanently installed in an ultra-high-vacuum (UHV) chamber, while the ``proximal'' end is brought outside through a home-made vacuum feedthrough \cite{Abraham2014}. The UHV chamber is a commercial glass cuvette epoxied to a Titanium flange and reaches a base pressure of $2\times 10^{-10}$ mbar, sufficient for many cold-atom-based platforms; for experiments using quantum degenerate gases it could be easily decreased by using an epoxy-free chamber and a bare or a metal-coated fiber. A pair of low-inductance magnetic coils and two frequency-stabilized \unit{780}{\nano\meter} external cavity diode lasers are used to create a conventional six-beam magneto-optical trap (MOT) for Rubidium 87 at \unit{3}{\milli\meter} from the distal end of the fiber. The MOT is loaded from a background vapor created with two Rb dispensers placed inside the DN40 vacuum tube leading to the cuvette. After polarization-gradient cooling, the cloud contains $2.2(5)\times 10^6$ atoms at \unit{40(5)}{\micro\kelvin} with a peak density of $1.5\times 10^{11}$ cm$^{-3}$.

\begin{figure}[tbp]
\centering
\includegraphics[width=85mm]{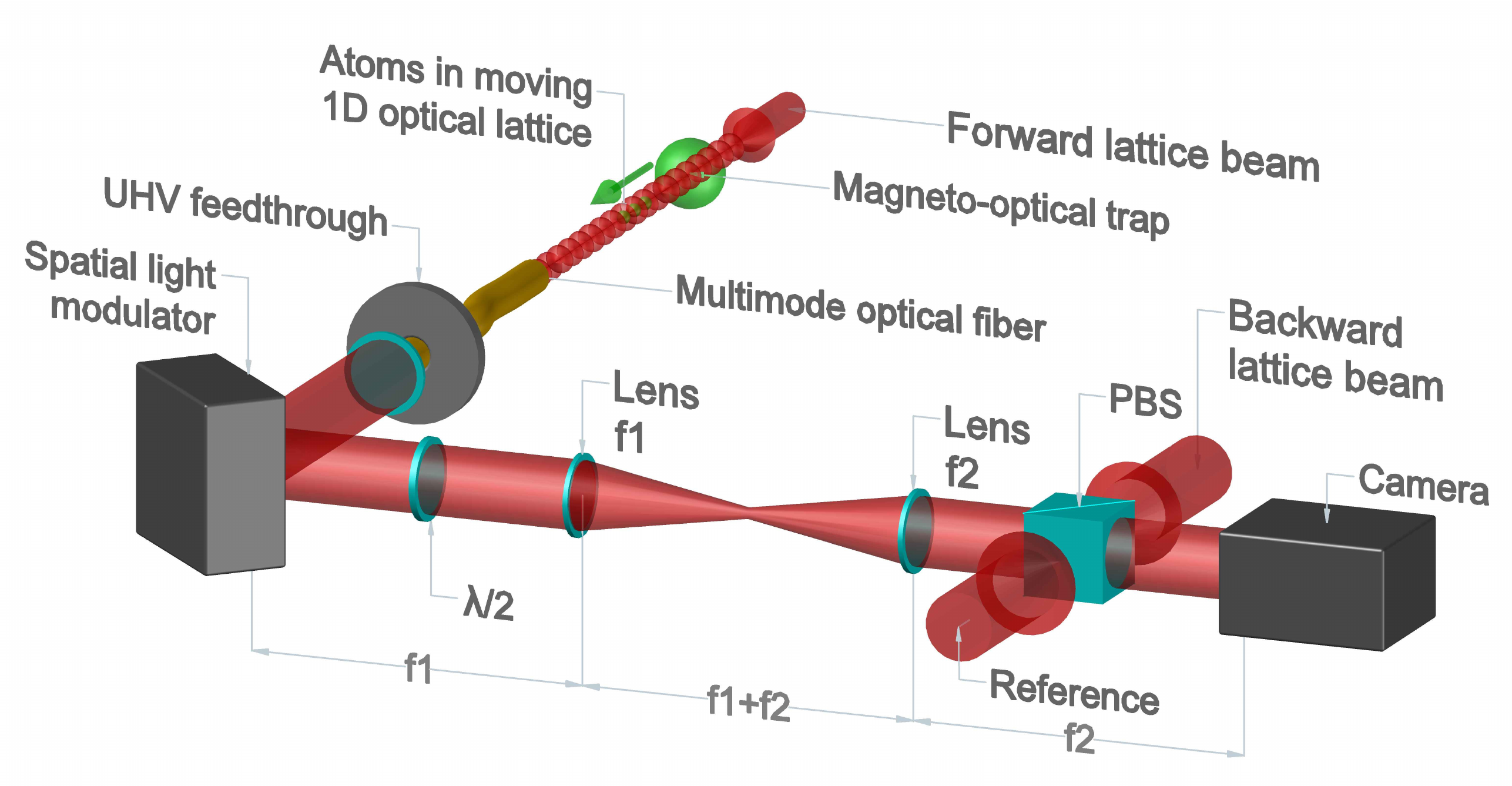}
\caption{Transport of cold atoms in a moving 1D optical lattice formed using phase-conjugated beams transmitted through a multimode fiber. The forward-propagating lattice beam is transmitted through the MMF, the resulting speckle field is measured, and a phase-conjugated backward-propagating lattice beam is formed using a spatial light modulator. The afocal telescope (two lenses with focals f1 and f2) simplifies the alignment and allows us to automatically include the SLM's aberrations in the phase-conjugation protocol (see text).}
\label{FigTransportSetup}
\end{figure}

All subsequent manipulations of the atoms rely on the use of adaptive optics at the proximal end of the MMF. In the experiments presented here, beams exiting the UHV system through the MMF are collimated, separated on a dichroic mirror, and sent to two identical digital optical phase conjugation (DOPC) setups operating respectively at \unit{800}{\nano\meter} for atomic transport and trapping (shown in Fig.\ref{FigTransportSetup}) and at \unit{780}{\nano\meter} for atomic cooling. The first ``recording'' stage of DOPC consists in injecting into the distal end of the MMF a forward-propagating beam with the desired properties. The speckle emerging from the proximal end interferes with a reference beam whose phase is scanned in $2\pi/3$ steps, allowing us to extract the speckle field from the interference patterns measured on a CMOS camera. In the second ``playback'' stage, we switch off the reference and switch on a backward-propagating beam, which we phase-conjugate with the forward-propagating one using a reflective phase-only liquid-crystal spatial light modulator (SLM, Hamamatsu X10468). This time-reversed beam is injected in the MMF and emerges from the distal end in the same transverse mode as the forward-propagating one, allowing us to perform a set of operations on the atoms described in the next sections.

Two known practical difficulties of DOPC are the pixel-to-pixel alignment of the camera with the SLM and the compensation of the SLM's aberrations, which correspond to an undesired phase mask added to the programmed one. 
We found a simple and efficient solution to both problems at once, shown in Fig.\ref{FigTransportSetup}. The SLM and the camera are installed in series and conjugated using an afocal telescope, which allows us to directly visualize the pixels of the SLM on the camera and to easily align the two devices. Moreover, as the forward-propagating beam experiences the SLM's aberrations as well, they are compensated by the phase-conjugation process together with the much stronger beam distortion introduced by the MMF. Finally, the use of a polarizing beamsplitter (PBS) and of a half-wave plate ($\lambda$/2) rotated by $45^\circ$ between the recording and the playback stages allows us to optimally use the power in the polarization controlled by the SLM. In practice, after imprinting the phase-conjugation mask on the SLM, we compensate residual alignment and focusing errors by adding a combination of the 3 lowest Zernike polynomials to the SLM's phase mask. We optimize the Zernike coefficients to maximize the overlap of the forward- and backward-propagating beams at the distal end of the MMF, by measuring the backward coupled power in the singlemode fiber emitting the forward-propagating beam. 

\section{Atomic transport}
\label{SectTransport}

\begin{figure}[tbp]
\centering
\includegraphics[width=85mm]{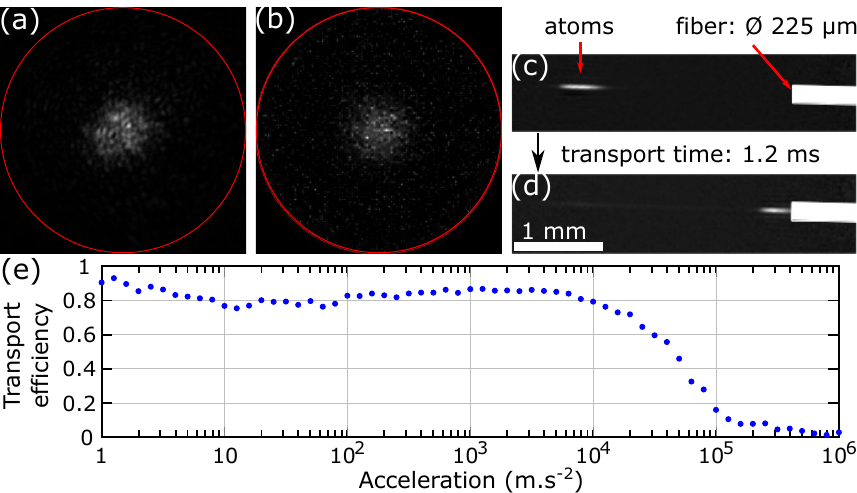}
\caption{(a): Measured intensity of the backward-propagating lattice beam emerging from the \unit{200}{\micro\meter} multimode fiber core (red circle). (b): Simulation of the measurement in (a) using the model described in the text. (c),(d): Absorption images of the atoms loaded in the 1D lattice, before (c) and after (d) transport. The white rectangle is the multimode fiber. (e): Transport efficiency as a function of acceleration for \unit{1}{\milli\meter} transport.}
\label{FigTransportData}
\end{figure}

Once the atoms are loaded in the MOT, we bring them closer to the distal end of the MMF in order to fully exploit its high numerical aperture. For this we use DOPC to create a moving 1D red-detuned far-off-resonant optical lattice. A forward-propagating \unit{100}{\milli\watt} beam at \unit{800}{\nano\meter} produced by a Ti:Sapphire laser is sent through the MOT and injected in the distal end of the fiber, the \unit{15}{\micro\meter} waist being located midway between the MOT and the MMF. The time-reversed beam emerges from the MMF and interferes with the forward-propagating beam to form an attractive standing-wave optical potential for the atoms. The $\sim 20\%$ overall efficiency of the conjugation process is mainly limited by: (i) the fact that we measure and phase conjugate only one polarization mode at the proximal side of the fiber ($60\%$), (ii) the SLM diffraction efficiency ($65\%$) determined by the SLM's ability to resolve $\sim$\unit{80}{\micro\meter} speckle grains with \unit{20}{\micro\meter} pixels, (iii) the efficiency of conjugation by phase-only modulation with the SLM ($\pi/4\simeq 79\%$ \cite{Cizmar2011}) and (iv) the losses on optical elements, in particular with the reflections from the MMFs end faces ($85\%$).
Therefore, while $\sim 20\%$ of the $100$-mW backward-propagating power is properly phase-conjugated creating a lattice of $1.5$ mK depth, $\sim 30\%$ of it is appears as a speckle field which interferes with the lattice and creates noise. Assuming an exponential distribution of the speckle intensities and a uniform distribution of speckle phases, we model consistently the experimentally measured intensity of the backward-propagating light exiting the MMF (Fig. \ref{FigTransportData}). This allows us to estimate the rms noise added to the lattice depth as $\sim 25\%$ of its ideal maximal value on the fiber face. However, this noise decreases with the distance to the fiber because the speckle strongly diverges whereas the conjugated mode remains collimated. Thus, this noise is expected to reduce the transport efficiency and increase the atomic-cloud temperature (due to a fluctuating trapping potential seen by the moving atoms) only over the last section of the transport when the atoms reach the high-resolution region of the MMF.

The atoms are transferred from the MOT into the 1D lattice by smoothly switching it on a the end of the MOT loading phase. The magnetic field gradient is then increased from 30 to 60 G/cm in 10 ms to compress the MOT, then switched off for 6 ms of polarisation-gradient cooling. At the end of this process, $\sim 1\times 10^5$ atoms at \unit{40(5)}{\micro\kelvin} are loaded in the lattice. The atoms are then accelerated  by linearly increasing the detuning between the forward- and backward-propagating beams, then decelerated in a similar way and come to a stop. As shown on Fig. \ref{FigTransportData}(e), the transport efficiency is nearly constant for accelerations up to \unit{10^4}{\meter .\second^{-2}}, which is consistent with the estimated spilling limit of \unit{3\times 10^5}{\meter .\second^{-2}} lowered by the transport-induced heating and the initial atomic temperature \cite{Schrader2001}. Additional losses appear next to the fiber due to atomic collisions with its surface. In practice, we form a cloud of  $\sim 6\times 10^4$ atoms at \unit{50}{\micro\kelvin} centered at $\sim$\unit{200}{\micro\meter} from the fiber's tip (see Fig. \ref{FigTransportData} (c) and (d)), where the optical resolution is maximal and limited by the numerical aperture of the fiber. Therefore, the phase-conjugated lattice potential allows a fast and efficient transport to this region.

\section{Optical microtraps}
\label{SectMicrotraps}

At this stage the atoms can be loaded into optical microtraps created using the MMF. In the ``recording'' stage of DOPC, these traps are formed using an auxiliary high-resolution optical system facing the distal end of the MMF, based on a digital micromirror device (DMD) illuminated with a Gaussian beam transmitted through a single-mode fiber. The DMD generates a programmable amplitude hologram, the first order of which is focused near the distal end of the MMF \cite{Zupancic2016}. By measuring this system's aberrations and applying a correction mask to the DMD \cite{Vitrant2019b} in addition to the desired hologram, this system becomes nearly diffraction-limited and allows us to focus light to \unit{1.2}{\micro\meter} waists near the distal end of the MMF. Arbitrary microtrap patterns can be generated, with no observable interferences between traps placed as close as \unit{1.5}{\micro\meter}.

\begin{figure}[tbp]
\centering
\includegraphics[width=85mm]{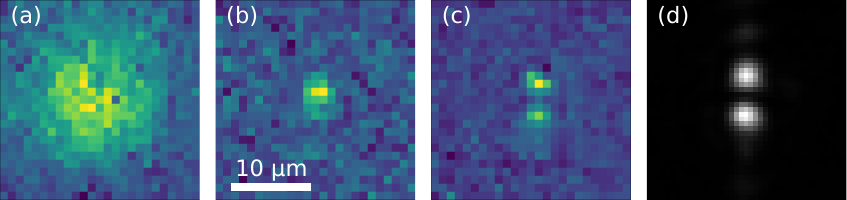}
\caption{ (a-c): Atoms loaded in (a) a small trap with a waist of \unit{5}{\micro\meter}, (b) a microtrap with a waist of \unit{1}{\micro\meter}, and (c) two such microtraps distant by \unit{5}{\micro\meter}, imaged through the MMF \cite{Vitrant2019b} with a \unit{100}{\micro\second} exposure and a $\times 100$ averaging.  As the transported cloud is rather dense, each microtrap is loaded with an atom number on the order of $10$. (d): Trapping light of (c) imaged using the DMD as a scanning-point detector (see text).}
\label{FigTrapData}
\end{figure}

Once the phase-conjugated counterparts of these beams are produced, the DMD-based source is switched off and the microtraps are generated via the multimode fiber only.
Their optical properties can be characterized by using the DMD-based source ``in reverse'' as a high-resolution scanning point detector, by focusing it to the desired spot and measuring how much backward-propagating light couples into its single-mode fiber (see Fig. \ref{FigTrapData}(d)). We cross-checked these measurements by characterizing a single MMF-generated microtrap with an independent NA=0.55 objective. Both approaches lead to measured waists of \unit{1.6}{\micro\meter}, consistent with the convolution of \unit{1.2}{\micro\meter} spots with the point spread functions of the characterization devices. Because the speckle produced by the microtraps at the proximal end of the MMF is approximately 2 times finer than for the 1D lattice, the diffraction efficiency of the SLM decreases leading to an overall DOPC efficiency of $\sim 2.5\%$. Nevertheless, this efficiency corresponds to a contrast of more than 600 between a \unit{1.2}{\micro\meter} trap formed at \unit{100}{\micro\meter} from the fiber's end, and the background speckle spread over a disk with a \unit{300}{\micro\meter} diameter. For \unit{100}{\milli\watt} of backward-propagating power this gives a depth of $-U/k_B \simeq 3.4$ mK for a single microtrap.

Both the lattice and the microtrap(s) use \unit{800}{\nano\meter} light, so we use a single DOPC setup to generate them both. We display them simultaneously because the dynamic response of the SLM to switch one hologram to the other is too slow relative to the timescale of the trapped-atom motion. The combination of the two holograms reduces by a factor of 0.4 the amount of backward-propagating light coupled to the lattice and to the microtrap, reducing their depths to $-U/k_B \simeq 1$ mK and $1.4$ mK, respectively. As the superposition of the two holograms is coherent, we optimize their relative phase factor to maximize the microtrap's depth. Once the atoms are brought next to the fiber, the forward-propagating lattice beam is linearly switched off in \unit{20}{\milli\second}. The backward-propagating lattice beam is then too weak to provide a sufficient longitudinal confinement, and atoms which were not loaded in the microtrap(s) leave the area in \unit{1}{\milli\second}. Figure \ref{FigTrapData} shows atoms loaded in traps of various geometries, absorption-imaged through the multimode fiber \cite{Vitrant2019b}.

\section{Cooling}
\label{SectCooling}
\FloatBarrier

\begin{figure}[tbp]
\centering
\includegraphics[width=85mm]{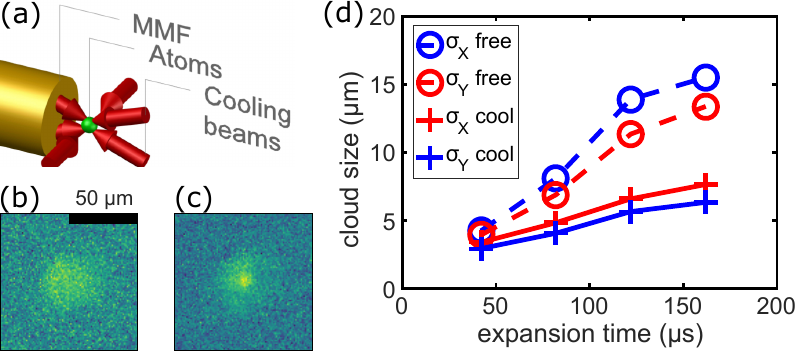}
\caption{(a): Doppler cooling of atoms after transport, using the multimode fiber to create two out of the six cooling beams. (b),(c): Absorption images of the atomic cloud after \unit{120}{\micro\second} of expansion, without (b) and with (c) cooling beams. (d): Expansion of the atomic cloud released from a small trap, without (dashed lines) and with (full lines) cooling beams.}
\label{FigCooling}
\end{figure}

We can use the phase-conjugation setup at \unit{780}{\nano\meter} to perform Doppler re-cooling using 3 pairs of counter-propagating beams (Fig.\ref{FigCooling}). Two of these beams, focused to \unit{100}{\micro\meter} waists, are injected in the distal end of the MMF with $\pm 20^\circ$ angles to the horizontal plane. Two counter-propagating beams emerging from the MMF are then created using DOPC at \unit{780}{nm}. The two remaining beams, with \unit{500}{\micro\meter} waists, are parallel to the MMFs end face. All six cooling beams are detuned \unit{12}{\mega\hertz} below the transition between the states $5S_{1/2},F=2$ and $5P_{3/2},F=3$ and set to $1.5$ times its saturation intensity. A weak beam repumps the atoms from the dark  state $5S_{1/2},F=1$.

To characterize the effect of the cooling beams, we switch off the trapping beams and measure the cloud's expansion with and without cooling. As shown on Fig. \ref{FigCooling}, in presence of cooling the expansion is significantly slower, allowing the cloud to maintain a higher density for a longer time. In turn, once the atoms are loaded in a microtrap, the near-resonant cooling beams lead to light-induced collisions that generate atomic-density-dependent losses. The MMF-generated microtraps are, in principle, confining enough to demonstrate sub-Poissonian atom statistics and, by decreasing the density of the transported cloud, reach the single-atom regime\cite{schlosser2001sub}. This regime could not be observed due to the low efficiency of our detection setup.

\section{Discussion}

Besides being much more compact, a multimode fiber used as a system for atomic manipulation has a few interesting features compared to bulk optics. In a traditional setup, creating beams with very different geometries such as those used for the 1D lattice, the cooling and the microtraps typically requires individual sets of optics. Here, thanks to the flexibility offered by adaptive optics and to the ability of the MMF to collect and guide light at the same time, all these functions are implemented using a single optical fiber, displacing the setup's complexity away from the UHV apparatus where space is the most scarce. The absence of a fixed working distance between the tip of the MMF and the atoms is an additional benefit of this approach.

On the other hand, the use of fibered optics constrains the wavelength of the trapping laser. For a given trap depth, one should increase the detuning from atomic resonances to reduce their off-resonant excitation. However, the corresponding increase in the optical power eventually heats the fiber and leads to thermal drifts that reduce the stability and efficiency of the DOPC after the characterization measurement. 

Compared to bulk optics, a multimode fiber offers a lower control over the polarisation and the spatial modes of the transmitted beams. At the current stage of our experiments, these problems have known solutions and can be considered as technical. The first issue can be solved by using two SLMs instead of one, controlling orthogonal polarisations. The second is, to a large extent, related to the limited resolution (792$\times$600 pixels) of our SLMs. Since 4160$\times$2464 resolutions are now available, the efficiency of DOPC can be significantly improved.

The key constrain of this approach, common to all systems based on adaptive optics, is the need for a ``guide star'' forming a well-defined optical pattern at the distal end of the fiber. In our case this guide star was created with bulk high-resolution optics facing the distal end of the fiber, which requires to keep a good optical access on one side of the atomic cloud. In some quantum platforms such as atomic chips this access exists and our approach can be used as-is. In others, where dense ($\rho\gtrsim 10^{14}$ cm$^{-3}$) clouds are manipulated and only a low-NA optical access is available, optical super-resolution techniques, based on the strong non-linearities of the atoms, can be used.  In systems where atoms are coupled to nanomechanical devices, a mechanical target can be envisioned.

\section{Conclusion}

We have shown that techniques developed for manipulating light in random optical media, originally with biomedical applications in mind \cite{Papadopoulos2013}, can be transposed to control cold atoms in spatially constrained quantum engineering platforms. Replacing bulk optics with a multimode high-NA optical fiber leads to a dramatic decrease in physical size and offers distinct features such as the possibility to form a variety of beam shapes which would otherwise require distinct sets of optics. Even though this approach offers a lower control over the spatial and polarisation modes of the light, it is sufficient to transport atoms in a moving 1D optical lattice, Doppler-cool them and load them in optical microtraps. With a reasonable experimental investment this degree of control can be significantly improved. More efforts are required to develop ``guide stars'' for the calibration of the fiber. It seems reasonable to assume that the solution to this problem will depend on the exact experimental setting.

\medskip
\noindent
\textbf{Disclosures}:
The authors declare no conflicts of interest.

\medskip
\noindent
\textbf{Funding Information}:
This work was funded by the DIM SIRTEQ project LECTRA, the IDEX grant ANR-10-IDEX-001-02-PSL PISE, and the ERC Starting Grant SEAQUEL.

\medskip
\noindent
\textbf{Acknowledgments}:
The authors thank P. Travers and  F. Moron for technical support, Q. Lavigne, T. Kouadou and J. Vaneecloo for their assistance at the early stage of the project, and S. Gigan for fruitful discussions.

\end{document}